\documentclass[conference]{IEEEtran}
\IEEEoverridecommandlockouts
\usepackage{cite}
\usepackage{amsmath,amssymb,amsfonts}
\usepackage{algorithmic}
\usepackage{graphicx}
\usepackage{textcomp}
\usepackage[utf8]{inputenc}
\usepackage[hyphens]{url}
\usepackage{balance}
\usepackage[numbers]{natbib}
\usepackage{multirow}
\usepackage{colortbl}
\usepackage{tabularx}
\usepackage{subcaption}
\usepackage{float}
\usepackage{xcolor}
\usepackage{fancyhdr} 
\usepackage{kantlipsum}
\fancypagestyle{plain}{ 
\fancyhf{} 
\fancyhead[C]{Conference on \LaTeX} 
 
} 
\usepackage{eso-pic} 

\def\BibTeX{{\rm B\kern-.05em{\sc i\kern-.025em b}\kern-.08em
    T\kern-.1667em\lower.7ex\hbox{E}\kern-.125emX}}
    
\begin{document}

\title{Inside the Right-Leaning Echo Chambers: Characterizing Gab, an Unmoderated Social System}

\author{\IEEEauthorblockN{Lucas Lima\IEEEauthorrefmark{1}, Julio C. S. Reis\IEEEauthorrefmark{1}, Philipe Melo\IEEEauthorrefmark{1}, Fabricio Murai\IEEEauthorrefmark{1},\\Leandro Araújo\IEEEauthorrefmark{1}, Pantelis Vikatos\IEEEauthorrefmark{2}, Fabrício Benevenuto\IEEEauthorrefmark{1}}
\IEEEauthorblockA{\IEEEauthorrefmark{1}Universidade Federal de Minas Gerais (UFMG), Brazil\\
\IEEEauthorrefmark{2}IIT-CNR, Pisa, Italy}
\{lucaslima, julio.reis, philipe, murai, leandroaraujo, fabricio\}@dcc.ufmg.br, p.vikatos@iit.cnr.it
}




\maketitle


\IEEEoverridecommandlockouts 
\IEEEpubid{\parbox{\columnwidth}{\vspace{8pt} 
\makebox[\columnwidth][t]{IEEE/ACM ASONAM 2018, August 28-31, 2018, Barcelona, Spain} 
\makebox[\columnwidth][t]{978-1-5386-6051-5/18/\$31.00~\copyright\space2018 IEEE} \hfill} \hspace{\columnsep}\makebox[\columnwidth]{}} 
\IEEEpubidadjcol 

\begin{abstract}

The moderation of content in many social media systems, such as Twitter and Facebook, motivated the emergence of a new social network system that promotes free speech, named Gab. Soon after that, Gab has been removed from Google Play Store for violating the company's hate speech policy and it has been rejected by Apple for similar reasons. In this paper we characterize Gab, aiming at understanding who are the users who joined it and what kind of content they share in this system. Our findings show that Gab is a very politically oriented system that hosts banned users from other social networks, some of them due to possible cases of hate speech and association with extremism. We provide the first measurement of news dissemination inside a right-leaning echo chamber, investigating a social media where readers are rarely exposed to content that cuts across ideological lines, but rather are fed with content that reinforces their current political or social views.
\end{abstract}


\section{Introduction} \label{sec:introduction}

Free speech and hate speech have opened a long debate between regulation and freedom of expression, especially in the online space. Currently, most online social networks have strict policies against hate speech and moderate the content posted by its users. In particular, Twitter, Facebook, Google (YouTube), and Reddit have largely increased removals of hate speech~\cite{reu-hate, Chandrasekharan:2017:YCS:3171581.3134666}. This scenario has motivated the emergence of a new social network system, named Gab\footnote{\url{https://gab.ai/}}, which was originally designed to be a social network that promotes free speech.

In spite of advocating liberty and freedom of speech, Gab has received several criticisms regarding the content shared there. The lack of moderation in this system started to attract users banned due to hate speech from other social networks~\cite{gab-banned}. Gab has also been pointed as a safe space for alt-right groups~\cite{bbc-gab}. This environment favors the emergence of clusters of like-minded individuals and the polarization of opinions, creating information bubbles around users who are fed with information that reinforces their current political or social views. This phenomenon is known as the ``echo chamber effect" and it has been extensively studied recently~\cite{shu2017fake, quattrociocchi2016echo}.

Several efforts have proposed ideas to mitigate the effect of ``echo chambers'' or ``filter bubbles'', either by introducing diversity in the news that users are consuming~\cite{munson2013encouraging,park2009newscube} or by highlighting posts that evoke similar reactions from opposite political views~\cite{babaei2018purple}. Despite the importance of all these efforts, little is still known about exactly what happens inside an echo chamber. This lack of understanding is likely due to the difficulty of splitting apart what is a bubble in social networks such as Facebook and Twitter. 

Therefore, understanding what happens inside Gab can provide us valuable insights on the study of right-leaning echo chambers. Our effort consists of answering the following research questions concerning Gab users and posts:

\vspace{0.1cm}
\noindent
\textbf{Research Question 1 (RQ1)}: Who are the users who joined Gab in terms of extremist views, political leaning and ability to spread news?


\vspace{0.1cm}
\noindent
\textbf{Research Question 2 (RQ2)}: What kind of content is shared in Gab? What are the news shared within this system?


\vspace{0.1cm}


Our findings show that Gab is a very politically oriented system, in agreement with concurrent work~\cite{zannettou2018gab}. Furthermore, we show that the majority of Gab users are conservative, male, and Caucasian. We were able to identify many users listed as extremists by the main media who showed to be influential and very active in the Gab network. The most popular type of discussion in Gab posts is focused on politics and conservatism. We also identified a few cases in which messages are considered highly toxic. 

We also show that, although most of the (unique) news domains present in shared URLs are considered left-leaning, conservative news outlets comprise a larger fraction of the shared links. By quantifying the popularity of sites whose links are disseminated inside this ideology-biased community, we show that popular news domains shared on Gab are not popular on Facebook or on the Internet. 


The rest of the paper is organized as follows. Section~\ref{sec:relatedWork} briefly reviews related work. Then, we present our datasets and the strategies we used to collect them in Section~\ref{sec:dataset}. Section~\ref{sec:results1} and Section~\ref{sec:results2} provide an in-depth investigation of our two research questions.  Section~\ref{sec:conclusion} provides our conclusions.  
\section{Related Work} \label{sec:relatedWork}

In this section, we review related work along two dimensions: (i) studies related to freedom of expression and (ii) efforts related to news sharing and propagation in social media.

\subsection{Freedom of Expression}

Freedom of speech is guaranteed by the first amendment in the USA and by other similar laws in liberal countries. A key argument pointed out by supporters of absolute free speech is that it is impossible to give power to anyone to restrict hate speech without allowing bans against non-hate speech. On the other hand, other theories defend that even free speech needs regulation. For instance, back in 1945, Karl Popper coined the paradox of tolerance~\cite{popper2012open} as \textit{Unlimited tolerance must lead to the disappearance of tolerance.} The key insight from these arguments is that if a society is tolerant without limit, their ability to be tolerant will eventually be seized or destroyed by the intolerants. More recently, Mitchell~\cite{mitchell2016liberalization} analyzes how freedom to speak is regulated in liberal states and points liberal states guarantees freedoms,  which require the creation of legal controls and regulations to ensure the social order.

These paradoxes and arguments, however, are still just theories, as there has been a lack of empirical data to support them. More important, many social systems have recently seen themselves in the midst of these paradoxes. As hate speech grows in the Web~\cite{silva2016@icwsm,mondal2017}, they needed to create hate speech policies and regulate what is expressed in their systems. The data within Gab now allow researchers to empirically measure what really happens in an unlimited free speech environment. Thus, by characterizing this system, we hope to bring insights that can contribute to this discussion.

\subsection{News Sharing and Propagation}

The dissemination of news in social networks has been the subject of several studies~\cite{bhattacharya2012sharing} focusing on different aspects, such as bias~\cite{chakraborty2016dissemination} and political news~\cite{an2014sharing}, or yet, the characteristics of players (or spreaders)~\cite{hu2012breaking}. In the case of Twitter, for example, Rodrigues \textit{et al.}~\cite{rodrigues2011word} show that retweets are responsible for increasing the audience of URLs by about two orders of magnitude. 

The characteristics (or factors) influencing news sharing in social media are explored by Lee and Ma~\cite{lee2012news}. Based on uses and gratifications, social cognitive theories, and interviews, the authors explore the influence of information seeking, socializing, entertainment, status seeking and prior social media sharing experience on news sharing intention. 
More recently, Garimella \textit{et al.}\cite{garimella2018political} study political echo chambers on social media. They examine the interaction between shared opinions and dissemination, dividing Twitter users in three categories: (i) users who are exposed to content that reinforces their opinions, (ii) users who try to bridge echo chambers by sharing content with diverse leaning, and (iii) gatekeepers, which are users who consume diverse content but produce content that fuels the bubbles. Differently, Reis \textit{et al.}~\cite{reis2017demographics} investigate the demographics of users in Twitter and how it affects news sharing. They show that white and male users tend to share more news on Twitter, biasing the news audience towards the interests of these demographic groups. 

Finally, in a very recent work, Ottoni \textit{et al.}~\cite{ottoni2018analyzing} investigate the presence of hateful content and discriminatory bias in right-wing YouTube channels, contributing to a better understanding of the behavior of right-wing YouTube users.



Overall, our study is complementary to previous efforts as our work provides an in-depth investigation of what kind of news are shared in a right-leaning echo chambers, thus providing a broader understanding of Gab users and content.

\section{Methodology} \label{sec:dataset}

In order to characterize Gab as a right-leaning echo chamber and to analyze its news sharing ecosystem, we describe next our strategy for data collection. We start by presenting an overview of the social network.

\subsection{The Gab Social Network}

In essence, Gab is very similar to Twitter, but barely moderates any of the content shared by its users. According to Gab guidelines, the site does not allow illegal pornography and promotion of violence and terrorism\footnote{\url{https://gab.ai/about/guidelines}}. Everything else towards free speech is allowed. Posts are limited to $3,000$ characters and there is an upvoting and downvoting system for posts, which does not exist on Twitter. Additionally, posts are categorized as News, Politics, Art etc., that users can use to find popular content on specific topics.

\subsection{Data Collection}

Our dataset includes data from users profiles, their lists of followers and friends, and all their messages. Each user in our dataset has an user ID, name, screen name, number of friends, followers, posts, account creation date, information of whether the profile is verified, PRO (subscribers), or Premium (users paid for their content), profile picture URL, and a profile bio. Each post comes along with a timestamp and its category. 

We crawled users and their posts by following a Breadth-First Search (BFS) scheme on the graph of followers and friends. We used as seeds users who authored posts listed by categories in the Gab main page. We implemented a distributed and parallel crawler that ran in August 2017 which took only three days to gather all users reachable from the seeds. In total, our dataset comprises $\textbf{171,920}$ users (the estimated number of users in August 2017 was $225$ thousand~\cite{mashable-gab}) and a total of $\textbf{12,829,976}$ posts. 

After collected, we preprocessed these data by discarding all messages marked as repost, quotes or unlabeled messages (i.e., messages not associated with any category). After
these steps, we reduced the data to a subset containing all the relevant information to our characterization: a sample of $\textbf{1,297,165}$ posts incorporating useful information for analyses discussed in the following sessions. 

\noindent{\textit{News Dataset.}} In order to provide a measurement of news dissemination inside Gab, we collected URLs' headlines of posts categorized as news containing URLs ($315,935$) according to the following two steps. First, we attempt to parse the HTMLs retrieved by the links and search for specific tags, such as \texttt{<title>}. When this is not possible, we match regular expressions to retrieve potential headlines of links from its own URL. Following these steps, 76.8\% out of the total of news messages have an associated headline. The following sections delve deeper in the study of these URLs, its domains, and headlines.

\section{RQ1: Who joined Gab?} \label{sec:results1}

In this section, we provide a series of analyses aiming at depicting who are the users that joined Gab. 

\subsection{When did users join Gab?}

Since the release of the service in August 2016, Gab has experienced a constant and slow increase in its number of users, except for a large peak around November 2016, during the US presidential election period. From all users we crawled, 51.4\% joined Gab from October to December, in 2016, 28.7\% of them just in November 2016. This suggests that the US presidential election represented an important motivation for users to join Gab. Zannettou {\it et al.}~\cite{zannettou2018gab} also highlight that Gab has become popular among users in the occurrence of political events such as the 2017 Charlottesville Unite the Right rally. 

Next, we attempt to infer the ideological leaning of Gab users.

\subsection{Political Leaning of users on Gab}

In order to measure the political bias of Gab users, we used a recently introduced framework~\cite{kulshrestha2017quantifying}, kindly shared by the authors. Their approach measures the political leaning of Twitter users, given a set of information from their friends. Thus, to identify the political leaning of Gab users, we first developed a strategy to identify the Twitter account of Gab users.   

\subsubsection{Identifying Gab users on Twitter}

We design a two-step methodology. First, we searched for the screen name of all Gab users in the \textit{Twitter REST API\footnote{\url{https://developer.twitter.com/en/docs}}}, identifying $62,291$ (36.23\%) users with exactly the same screen name in both systems. Then, we compare the profile name of these users from Gab with their respective profile names in Twitter, keeping only those for which we found an exact match. Our final dataset contains $23,030$ users, which corresponds to about 13\% of the total number of users.

\subsubsection{Inferring Political Bias of Twitter users} 

For each user identified as being the same on Twitter and Gab, we further gathered her lists of Twitter friends from the Twitter API. Then, the framework used to measure these users' political leaning~\cite{kulshrestha2017quantifying} was able to identify the leaning of $16,804$ users (73\% of the matched pairs). 

\begin{figure}[tb]
 \centering
\includegraphics[scale=0.55]{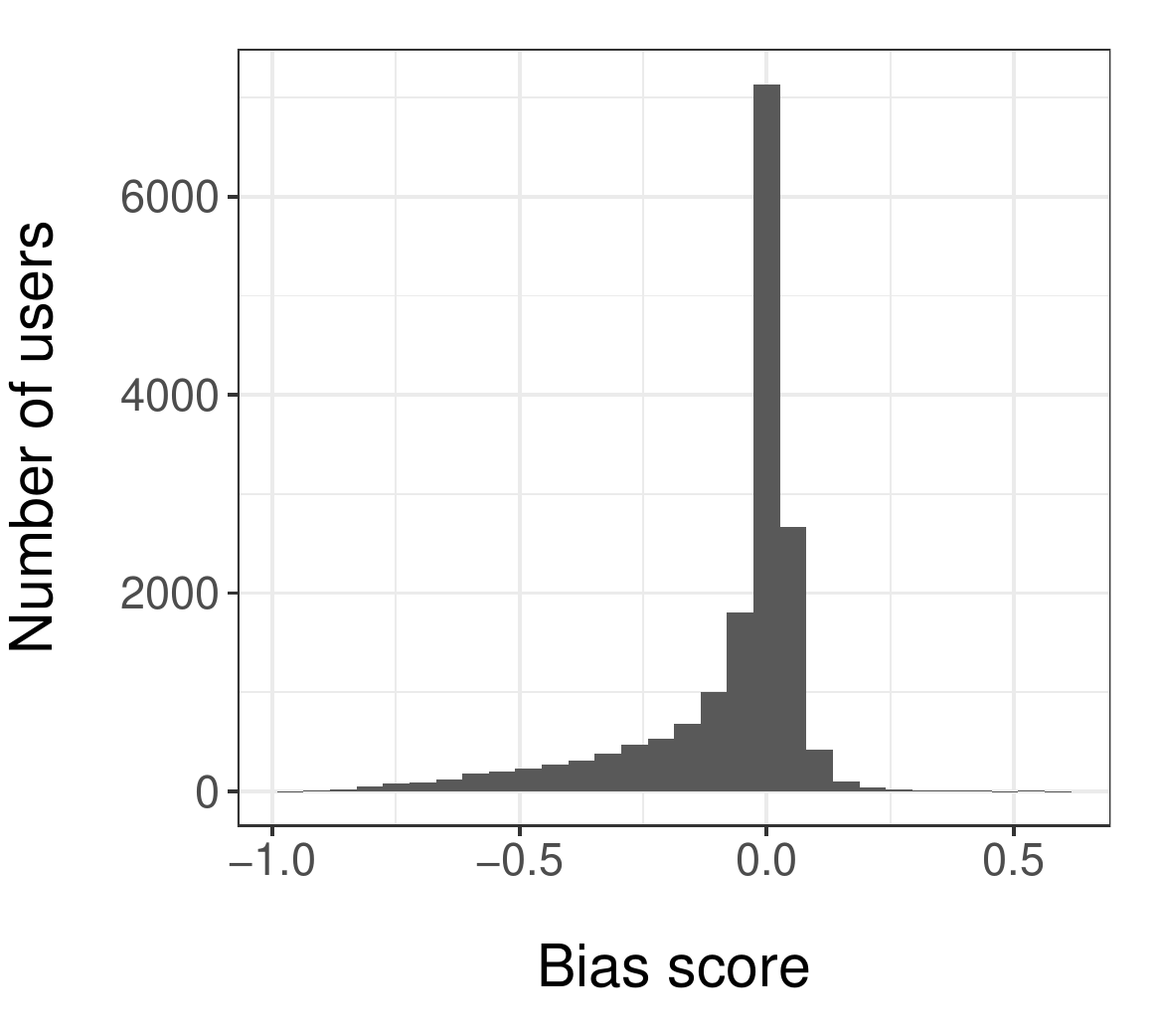}
\caption{\textbf{Distribution of bias scores for mapped users.}}\vspace{-15px}
\label{fig:biasDistribution}
\end{figure}

Figure~\ref{fig:biasDistribution} shows the distribution of bias scores. The closer the bias score is to~+1, the more liberal a user is. Users with bias score between -0.03 and +0.03 are considered neutral or moderate. We note a large number of extremely conservative users (i.e. those with scores close to~-1) in Gab and most of the liberal ones have ideological leaning scores quite close to 0 (i.e. moderate). Out of the $16,804$ users we sampled, $6,237$~(37.1\%) were inferred to be conservative, $2,925$~(17.4\%) as liberal, and $7,642$~(45.5\%) as moderate. Hence, the ratio between conservatives and liberals is about 2.13.

\vspace{0.1cm}

In order to contrast these values with those from other social networks, we use the Facebook audience API, following the same methodology of recent efforts~\cite{speicher-2018-targeted}. Facebook provides this API for advertisers to estimate the amount of users who are likely to match an advertising criteria\footnote{\url{developers.facebook.com/docs/marketing-api/audiences-api}}. We then used the Audience API to gather the political leaning of all Facebook users in the US. Overall, the inferred fraction of conservatives in Facebook is 33.6\%, 42.6\% of liberals and 23.8\% of moderates. The ratio between conservatives and liberals is about 0.79. Therefore, in comparison with Facebook, Gab has a much more conservative than liberal crowd.


One limitation of this analysis is that it requires us to match Gab and Twitter accounts as a first step, thus reducing the set of suitable users to 13.4\%. This is due to the fact the framework provided by Kulshrestha \textit{et al.}~\cite{kulshrestha2017quantifying} is a Twitter-based approach. We leave as future work the creation of a new inference framework based on the Gab network only.


\subsection{Gender and Race of Gab users}
In order to better understand demographic aspects of users in Gab, we 
used a methodology recently explored in previous efforts \cite{messias2017white,chakraborty2017@icwsm} to infer the gender and race of Gab users. The strategy consists of gathering the profile picture Web link of Gab users and submit these links into the Face++ API. Face++ is a face recognition platform based on deep learning which is able to identify gender (i.e.\ male and female) and race (limited to Asian, Black, and White) from recognized faces in images. We have discarded users whose profile does not have a face picture or does not have a recognizable face according to Face++. 
A recent work~\cite{chakraborty2017@icwsm} evaluated the effectiveness of the inference made by Face++, using human annotators to label randomly selected profile images from Twitter. They measured the
inter-annotator agreement in terms of the Fleiss' K score which was
1.0 and 0.865 for gender and race, respectively.

From the total number of Gab users we crawled, less than a half (47.22\%) have a profile picture Web link. Given these $82,215$ users, only $35,493$ have valid profile pictures that we were able to collect demographic information. 
Table~\ref{table:demographicsGab} reports the demographic distribution of the $35,493$ users in our dataset across the different demographic groups. We observe a prevalence of males (67.2\%) in comparison to females (32.7\%) and a high predominance of Whites (76.1\%) in comparison to Asians (15.8\%) and Blacks (8.2\%). This means that if we pick users randomly in our dataset, we would expect demographic groups with these proportions. These proportions, in particular the ones regarding race, differ from Facebook, where the White men correspond to 29.35\% of the total population~\cite{speicher-2018-targeted}.

\begin{table}[tb]
\centering
\resizebox{\columnwidth}{!}{
\begin{tabular}{|l|r|r|r|}
\hline
\multicolumn{1}{|c|}{\multirow{2}{*}{\textbf{Race}}} & \multicolumn{2}{c|}{\textbf{Gender}}                                      & \multicolumn{1}{c|}{\multirow{2}{*}{\textbf{Total}}} \\ \cline{2-3}
\multicolumn{1}{|c|}{}                               & \multicolumn{1}{c|}{\textbf{Male}} & \multicolumn{1}{c|}{\textbf{Female}} & \multicolumn{1}{c|}{}                                \\ \hline
Asian                                                & $3,676$ (10.4\%)                   & $1,920$ (5.4\%)                      & $5,596$ (15.8\%)                                     \\ \hline
Black                                                & $2,106$ (5.9\%)                    & 787 (2.2\%)                          & $2,893$ (8.2\%)                                      \\ \hline
White                                                & $18,078$ (50.9\%)                  & $8,926$ (25.1\%)                     & $27,004$ (76.1\%)                                    \\ \hline
\textbf{Total}                                       & $23,860$ (67.2\%)                  & $11,633$ (32.7\%)                    & $35,493$ (100\%)                                     \\ \hline
\end{tabular}
}
\caption{\textbf{Demographic distribution of nearly 36 thousand Gab users.}}\vspace{-15px}
\label{table:demographicsGab}
\end{table}

\subsection{Extremism in Gab} 

Next, we want to analyze whether far-right groups are present in Gab. According to the  SPLC\footnote{Southern Poverty Law Center \url{https://www.splcenter.org/}}, which is an American nonprofit legal advocacy organization specializing in civil rights dedicated to fighting hate, extremists in the United States come in different forms and follow a wide range of ideologies. As we look through the individual profiles of extremists provided by the SPLC~\cite{splc18:extremist} and the Anti-Defamation League (ADL)~\cite{adl18:naming}, we notice the presence of a few members of these movements in the platform. Table~\ref{table:usersADL} presents the list of 29 listed extremist users by the two sources with Gab accounts and their number of Gab followers as observed in August 2017. In particular, most people listed by the ADL (61.11\%) have a Gab account, reinforcing that this social network has been attracting a large number of extremely conservative users. 


\begin{table}[tb]
\centering
\resizebox{\columnwidth}{!}{
\begin{tabular}{llr}
\hline
\multicolumn{1}{|c|}{\textbf{Screen name}}    & \multicolumn{1}{c|}{\textbf{User name}}      & \multicolumn{1}{c|}{\textbf{\# of followers}} \\ \hline
\multicolumn{1}{|l|}{RealAlexJones}            & \multicolumn{1}{l|}{Alex Jones**}      	   & \multicolumn{1}{r|}{$14,962$}                 \\ \hline
\multicolumn{1}{|l|}{Alex\_Linder}              & \multicolumn{1}{l|}{Alex Linder}       	   & \multicolumn{1}{r|}{$963$}                 \\ \hline
\multicolumn{1}{|l|}{AndrewAnglin}            & \multicolumn{1}{l|}{Andrew Anglin**}         & \multicolumn{1}{r|}{$4,853$}                 \\ \hline
\multicolumn{1}{|l|}{AndyNowicki}             & \multicolumn{1}{l|}{Andy Nowicki}            & \multicolumn{1}{r|}{$75$}                    \\ \hline
\multicolumn{1}{|l|}{thewizardofthorntonpark} & \multicolumn{1}{l|}{Augustus Invictus}       & \multicolumn{1}{r|}{$128$}                    \\ \hline
\multicolumn{1}{|l|}{BillyRoper}               & \multicolumn{1}{l|}{Billy Roper}       	   & \multicolumn{1}{r|}{$407$}                 \\ \hline
\multicolumn{1}{|l|}{occdissent}              & \multicolumn{1}{l|}{Brad Griffin}            & \multicolumn{1}{r|}{$534$}                  \\ \hline
\multicolumn{1}{|l|}{BrittPettibone}          & \multicolumn{1}{l|}{Brittany Pettibone*}     & \multicolumn{1}{r|}{$23,335$}                 \\ \hline
\multicolumn{1}{|l|}{Cantwell}                & \multicolumn{1}{l|}{Christopher Cantwell***} & \multicolumn{1}{r|}{$2,728$}                  \\ \hline
\multicolumn{1}{|l|}{DanielFriberg}           & \multicolumn{1}{l|}{Daniel Friberg}          & \multicolumn{1}{r|}{$299$}                    \\ \hline
\multicolumn{1}{|l|}{RealDavidDuke}            & \multicolumn{1}{l|}{David Duke****}           & \multicolumn{1}{r|}{$982$}                 \\ \hline
\multicolumn{1}{|l|}{GavinMcInnes}           & \multicolumn{1}{l|}{Gavin McInnes}            & \multicolumn{1}{r|}{$2,681$}                  \\ \hline
\multicolumn{1}{|l|}{Posobiec}                & \multicolumn{1}{l|}{Jack Posobiec*}          & \multicolumn{1}{r|}{$2,944$}                  \\ \hline
\multicolumn{1}{|l|}{jartaylor}               & \multicolumn{1}{l|}{Jared Taylor*}           & \multicolumn{1}{r|}{$147$}                  \\ \hline
\multicolumn{1}{|l|}{TheMadDimension}         & \multicolumn{1}{l|}{Jason Kessler*}          & \multicolumn{1}{r|}{$454$}                  \\ \hline
\multicolumn{1}{|l|}{mattforney}              & \multicolumn{1}{l|}{Matt Forney**}           & \multicolumn{1}{r|}{$4,322$}                  \\ \hline
\multicolumn{1}{|l|}{matthewheimbach}         & \multicolumn{1}{l|}{Matthew Heimbach}        & \multicolumn{1}{r|}{$43$}                     \\ \hline
\multicolumn{1}{|l|}{mattparrott}             & \multicolumn{1}{l|}{Matthew Parrott}         & \multicolumn{1}{r|}{$56$}                    \\ \hline
\multicolumn{1}{|l|}{Cernovich}               & \multicolumn{1}{l|}{Mike Cernovich*}         & \multicolumn{1}{r|}{$27,462$}                 \\ \hline
\multicolumn{1}{|l|}{mikeenoch}               & \multicolumn{1}{l|}{Mike Peinovich}          & \multicolumn{1}{r|}{$703$}                  \\ \hline
\multicolumn{1}{|l|}{m}                       & \multicolumn{1}{l|}{Milo Yiannopoulos*}      & \multicolumn{1}{r|}{$39,891$}                 \\ \hline
\multicolumn{1}{|l|}{Pamela}                   & \multicolumn{1}{l|}{Pamela Geller*}           & \multicolumn{1}{r|}{$3,238$}                 \\ \hline
\multicolumn{1}{|l|}{ramzpaul}                 & \multicolumn{1}{l|}{Paul Ray Ramsey}          & \multicolumn{1}{r|}{$2,951$}                 \\ \hline
\multicolumn{1}{|l|}{pax}                     & \multicolumn{1}{l|}{Pax Dickinson}           & \multicolumn{1}{r|}{$12,218$}                 \\ \hline
\multicolumn{1}{|l|}{Richardbspencer}         & \multicolumn{1}{l|}{Richard B Spencer}       & \multicolumn{1}{r|}{$6,833$}                 \\ \hline
\multicolumn{1}{|l|}{RobertSpencer}            & \multicolumn{1}{l|}{Robert Spencer}           & \multicolumn{1}{r|}{$1,317$}                 \\ \hline
\multicolumn{1}{|l|}{TaraMcCarthy}            & \multicolumn{1}{l|}{Tara McCarthy***}        & \multicolumn{1}{r|}{$5,565$}                  \\ \hline
\multicolumn{1}{|l|}{voxday}                  & \multicolumn{1}{l|}{Theodore Beale*}         & \multicolumn{1}{r|}{$18,551$}                 \\ \hline
\multicolumn{1}{|l|}{Microchimp}              & \multicolumn{1}{l|}{Tim Gionet}              & \multicolumn{1}{r|}{$4$}                     \\ \hline
\multicolumn{3}{l}{\textit{* verified; ** verified and PRO; *** verified, PRO and premium;}} \\ 
\multicolumn{3}{l}{\textit{**** PRO}}
\end{tabular}}
\caption{\textbf{List of profiles considered extremist by the SPLC and ADL who were found on Gab.}}\vspace{-15px}
\label{table:usersADL}
\end{table}


A key question that arises about these listed extremist users in Gab is whether they are widely followed and whether they are active users in the system. 
Among the top 10 most followed users, four are extremist users according to the SPLC and ADL, and all of them have verified accounts (that is a form of veryfying identities). These considered extremist users, namely Milo Yiannopoulos (@m), Mike Cernovich (@Cernovich), Brittany Pettibone (@BrittPettibone) and Vox Day (@voxday), have on average $27,309.75$ Gab followers. The average number of followers over the list of 29 users is $6,160.2$, two orders of magnitude larger than the average number of followers for all Gab users we crawled (72.23). This means that posts of listed extremist users can reach a large number of different users within one hop. Approximately 35\% of Gab users we crawled follow at least one of these 29 extremist users.

In regards to the number of posts, we also note that these users share in average more posts than the average Gab user. The average number of Gab posts over all users is 55.69, whereas the average for listed extremist users is 571.66. This indicates that these users are not only more followed, but they are also more active in the system. 

\subsection{News spreaders}

We now analyze the profile of news spreaders in our dataset, i.e.\ people who share news sources within the news category in Gab posts. It is interesting to note that the majority of the active users within the news category (60.14\%) have posted at least one URL and, from these, 62.71\% have posted more than one URL.

Table~\ref{table:newsSpreaders} shows the top 10 news spreaders and their total number of posts as of August 2017. These users have shared on average $10,838.5$ posts. The user \textit{Constitutional Drunk}, associated with the right-biased USSA News website, had the largest number of posts, $59,378$. In fact, most publications from this user share content of that website, which happens to be the most shared domain. Only 37.9\% of the users listed as extremists have posts categorized as news.

Aiming at understanding the influence of these users in Gab, we also analyze their number of followers. While the average number of followers among all crawled users is low (72.23), the top 10 news spreaders have on average $4,128.8$ followers, two orders of magnitude higher than the overall mean. In total, $20,176$ out of $171,920$ Gab users follow at least one of these news spreaders, which implies that posts from the top 10 news spreaders can reach 11.74\% of the the social network users.

\begin{table}[tb]
\centering
\begin{tabular}{llr}
\hline
\multicolumn{1}{|c|}{\textbf{Screen name}}     & \multicolumn{1}{c|}{\textbf{User name}}       & \multicolumn{1}{c|}{\textbf{\# of posts}} \\ \hline
\multicolumn{1}{|l|}{USSANews}            & \multicolumn{1}{l|}{Constitutional Drunk}      	   & \multicolumn{1}{r|}{$59,378$}                 \\ \hline
\multicolumn{1}{|l|}{Zlatford}              & \multicolumn{1}{l|}{Zak}       	   & \multicolumn{1}{r|}{$13,388$}                 \\ \hline
\multicolumn{1}{|l|}{wrath0fkhan}             & \multicolumn{1}{l|}{wrath 0fkhan}          & \multicolumn{1}{r|}{$8,189$}                 \\ \hline
\multicolumn{1}{|l|}{histanvan}               & \multicolumn{1}{l|}{Harry2}       	   & \multicolumn{1}{r|}{$6,459$}                 \\ \hline
\multicolumn{1}{|l|}{Kek\_Magician}                 & \multicolumn{1}{l|}{Kek\_Magician}   & \multicolumn{1}{r|}{$4,248$}                 \\ \hline
\multicolumn{1}{|l|}{Lakeem}            & \multicolumn{1}{l|}{Lakeem Khodra***}           & \multicolumn{1}{r|}{$4,016$}                 \\ \hline
\multicolumn{1}{|l|}{Arwen777}                & \multicolumn{1}{l|}{Dani}            & \multicolumn{1}{r|}{$3,913$}                 \\ \hline
\multicolumn{1}{|l|}{weeklyflyer}          & \multicolumn{1}{l|}{Jerry}         & \multicolumn{1}{r|}{$3,213$}                 \\ \hline
\multicolumn{1}{|l|}{rabite}                   & \multicolumn{1}{l|}{Stankpipe}           & \multicolumn{1}{r|}{$2,925$}                 \\ \hline
\multicolumn{1}{|l|}{OpenQuotes}                 & \multicolumn{1}{l|}{OpenQuotes}          & \multicolumn{1}{r|}{$2,656$}                 \\ \hline
\multicolumn{3}{l}{\textit{*** verified, PRO and premium.}}
 
\end{tabular}
\caption{\textbf{Top 10 news spreaders.}} 
\label{table:newsSpreaders}
\end{table}
\section{RQ2: What users share in Gab?} \label{sec:results2}

In this section, we analyze the content shared in Gab. We start our analysis by characterizing typical language usage.

\subsection{Popular Words and Topics in Gab}

\begin{figure}[t]
\centering
\includegraphics[scale=0.4]{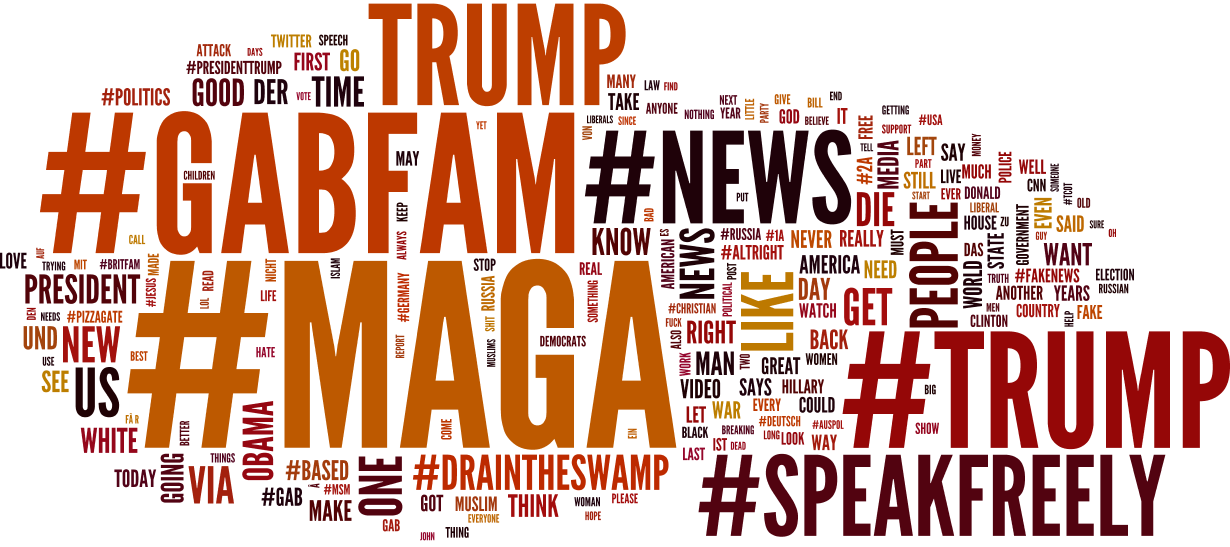}
\caption{\textbf{Word clouds for Gab posts.}}
\label{fig:wordcloud}
\end{figure}

Figure~\ref{fig:wordcloud} illustrates through one word cloud\footnote{\url{http://www.wordle.net/}} frequent terms of Gab posts for all posts from our dataset. Overall, this word cloud shows a strong political emphasis in Gab posts, with the existence of several terms which have been used by conservative political campaigns and their supporters. In particular, the hashtags \textit{MAGA}, which stands for \textit{M}ake \textit{A}merica \textit{G}reat \textit{A}gain, and \textit{DRAINTHESWAMP} were popularly mentioned as part of the Donald Trump's 2016 presidential campaign.

The analysis of posts' categories reinforces the strong political trait of the Gab social network. The category with the largest number of posts is News (35.74\%), usually associated with politics, followed by the categories Politics, Humor, AMA (Ask Me Anything) and Entertainment. The other categories -- Music, Technology, Art, Sports, Faith, Philosophy, Photography, Science, Finance and Cuisine -- sum up to $225,611$ (17.39\%) posts.

\begin{figure}[t]
 \centering
\includegraphics[scale=0.7]{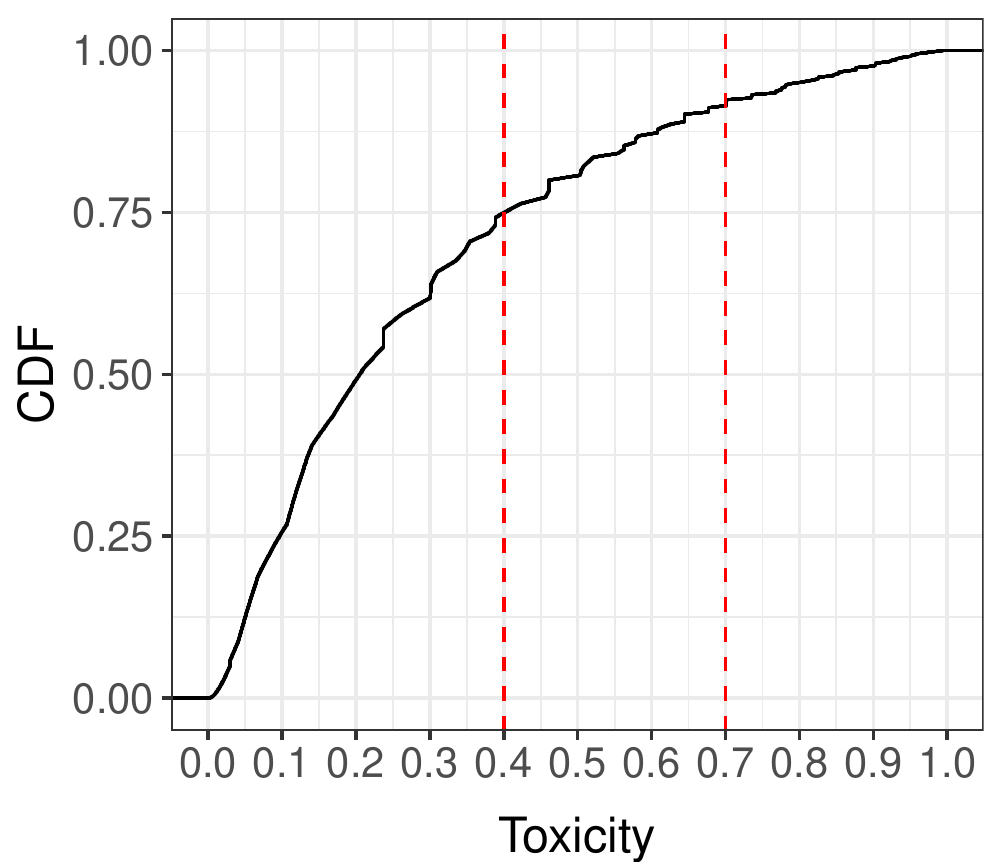}
\caption{\textbf{Cumulative distribution function (CDF) for toxicity scores of categorized posts.}}\vspace{-10px}
\label{fig:cdf}
\end{figure}

\subsection{Evidence of Hate Speech in Gab}

Recently, there have been several efforts towards identifying online hate speech, as the problem manifests itself in different systems and under different forms. A recent API  called Perspective\footnote{\url{https://www.perspectiveapi.com/}}, created by Jigsaw and Google's Counter Abuse Technology team, uses machine learning models to identify whether comments could be perceived as toxic or not, providing toxicity scores which ranges from 0 to 1. A toxic comment is defined by the creators of the project as \textit{a rude, disrespectful, or unreasonable comment that is likely to make you leave a discussion}. Given that properly detecting all the hate speech in Gab represents a methodological challenge as current methods for that are still very incipient, we provide evidence of hate speech in Gab through the analysis of Gab categorized posts' toxic scores.

Figure~\ref{fig:cdf} shows the cumulative distribution function (CDF) for the sample of $982,875$ Gab posts (75.77\% of the total number of categorized posts) from which we were able to gather toxicity scores from the API. Toxicity scores lower than 0.4 indicates posts unlikely to be perceived as toxic, while scores higher than 0.7 indicates posts likely to be perceived as toxic. Toxicity scores between 0.4 and 0.7 indicates unsureness whether posts will be perceived as toxic. We notice that more than 90\% of these posts (91.4\%) have toxicity scores less than 0.70, i.e.\ a large fraction of the total number of posts are either unlikely to be perceived as toxic, or may or may not be perceived as toxic. There are only $23,188$ posts (2.4\%) with toxicity score higher than 0.9.

Even though there is just a small percentage of Gab posts perceived as toxic, some posts with scores closer to 1 clearly incorporate hate speech towards groups of people of different religion and ethnic origin, race, sexual orientation and gender, e.g., \textit{I F*CKING\footnote{Wherever present, the ``*'' has been inserted by us, in order to lessen the impact of the offense words may inflict on some people, and was not part of the original text.} HATE N*GGERS THEY ARE ALL DISGUSTING AND THEY SHOULD ALL DIE A BUNCH OF F*CKING C*NTS} and \textit{F*ck Islam! Muhammad was a f*ggot pedophile! F*ck Islam! F*ck all of it! Your religion s*cks, your culture s*cks, your women s*ck, every f*cking thing about Islam is inferior, barbaric, and gay. F*CK ISLAM}. Therefore, we notice there are users who abuse the lack of moderation to disseminate hate.


\subsection{What kind of news are shared in Gab?}
Inferring bias of news sources often relies on strategies which combine analyses of the audience of news outlets and inspection of the published content. Recent research works such as Budak \textit{et al.}~\cite{budak2016fair}, Mitchell \textit{et al.}~\cite{mitchell2014political}, Bakshy \textit{et al.}~\cite{bakshy2015exposure}, and AllSides\footnote{https://allsides.com/media-bias/media-bias-ratings} infer the political leaning of over 600 news outlets using these techniques. Next, we briefly discuss each of these studies and, based on them, we analyze the presence of biased news sources on Gab.

Following a survey-based approach, Mitchell \textit{et al.}~\cite{mitchell2014political} classify the audience of popular news media outlets based on a ten question survey covering a range of issues like homosexuality, immigration, economic policy, and the role of government, classifying the political leaning of the audience in five categories: consistently liberal, mostly liberal, mixed, mostly conservative, and consistently conservative. On the other hand, Bakshy et al.~\cite{bakshy2015exposure} follow a news-based approach in which the authors derive the alignment score of media outlets by first identifying the political leaning of over 10 million Facebook users based on self-declarations, and then considering how users with different political leanings shared the stories published by these outlets. 

Other studies include approaches based on (a) content and (b) crowdsourcing. The former refers to the research of Budak et al.~\cite{budak2016fair} where the authors use a content-based approach to identify the slant of the top 13 U.S.\ news outlets and two popular political blogs. Regarding the latter, the website \url{AllSides.com} infers bias by encouraging its users to rate different news outlets in one of the five categories: left, lean left, center, lean right, and right.

Table~\ref{tab:dataset-political-bias_match} presents, for each of the aforementioned datasets, the number of news outlets inferred as Republican, Democrat, and Neutral, which were also shared on Gab posts categorized as news. We notice that regardless of the bias of news outlets, most domains are in fact found in Gab posts (e.g., 100\% of the news outlets with political leaning inferred by Budak et al. have been shared in Gab). However, we show next that news sources inferred as Republican are shared on average more often than those inferred as Democrat and Neutral.

\begin{table}[t] \centering 
\vspace*{2mm}
 \small
 \resizebox{\columnwidth}{!}{
 \begin{tabular} {|l|r|r|r|}
 \hline
  \textbf{Dataset} & \textbf{Republican} \textbf{(\%)}  & \textbf{Democrat} \textbf{(\%)} & \textbf{Neutral} \textbf{(\%)}  \cr \hline
Budak \textit{et al.} & 4 of 4 (100.00)  & 11 of 11 (100.00) & -   \cr \hline		
Mitchell \textit{et al.} & 6 of 7 (85.71) & 19 of 23 (82.61) & 1 of 3 (33.33) \cr 	\hline
Bakshy \textit{et al.} & 165 of 205 (80.49) & 203 of 260 (78.08) & -	\cr \hline	
AllSides & 29 of 40 (72.50)  & 26 of 40 (65.00) & 24 of 32 (75.00)  \cr \hline		
 \end{tabular} 
 }
 \caption{\textbf{Number of domains found in Gab posts which are categorized as news and coexistence in each dataset.}}\label{tab:dataset-political-bias_match}\vspace{-15px}
 \end{table}
 
\begin{figure*}[tb]
 
\begin{subfigure}{0.24\textwidth}
\includegraphics[width=0.9\linewidth]{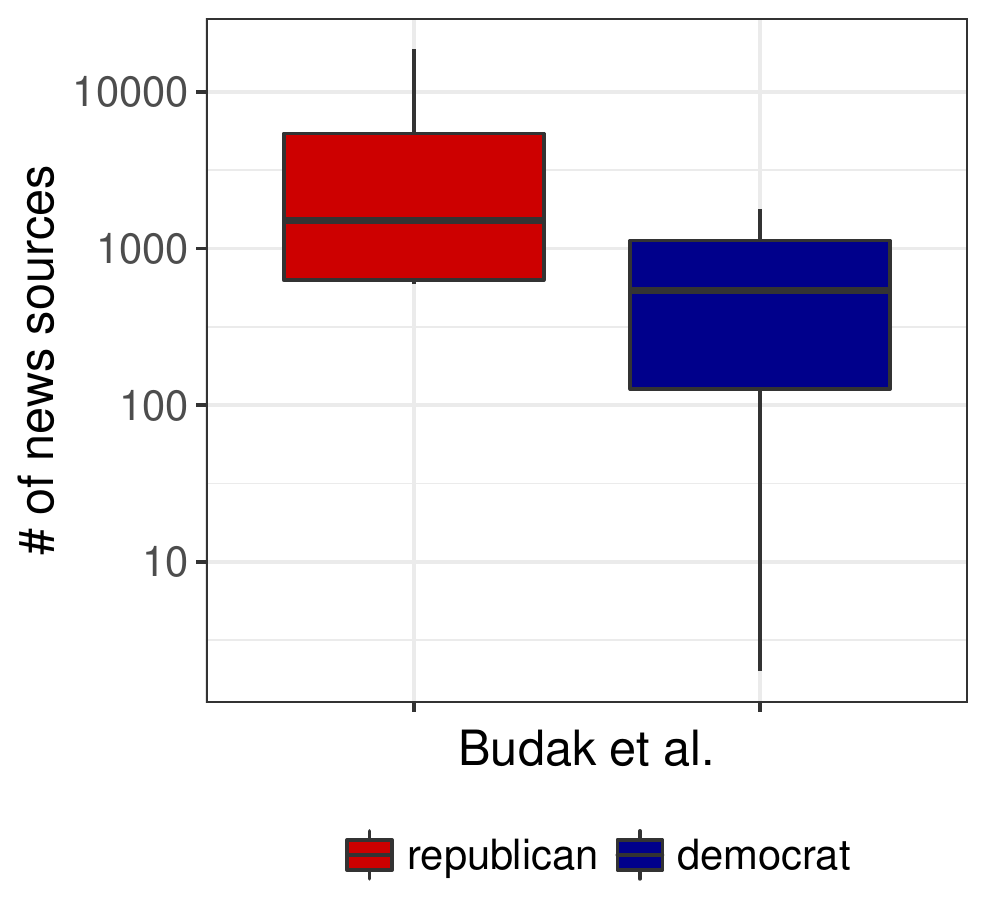} 
\caption{}
\label{fig:Budak}
\end{subfigure}
\begin{subfigure}{0.24\textwidth}
\includegraphics[width=0.9\linewidth]{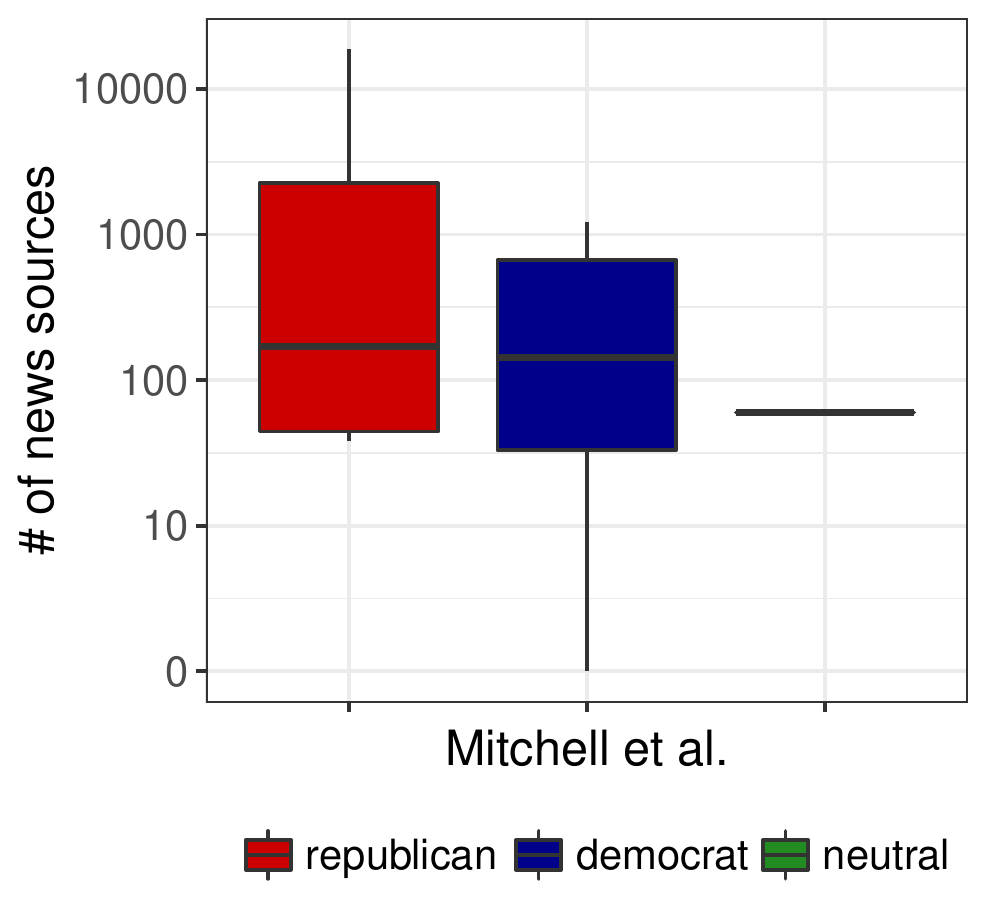} 
\caption{}
\label{fig:Mitchell}
\end{subfigure}
\begin{subfigure}{0.24\textwidth}
\includegraphics[width=0.9\linewidth]{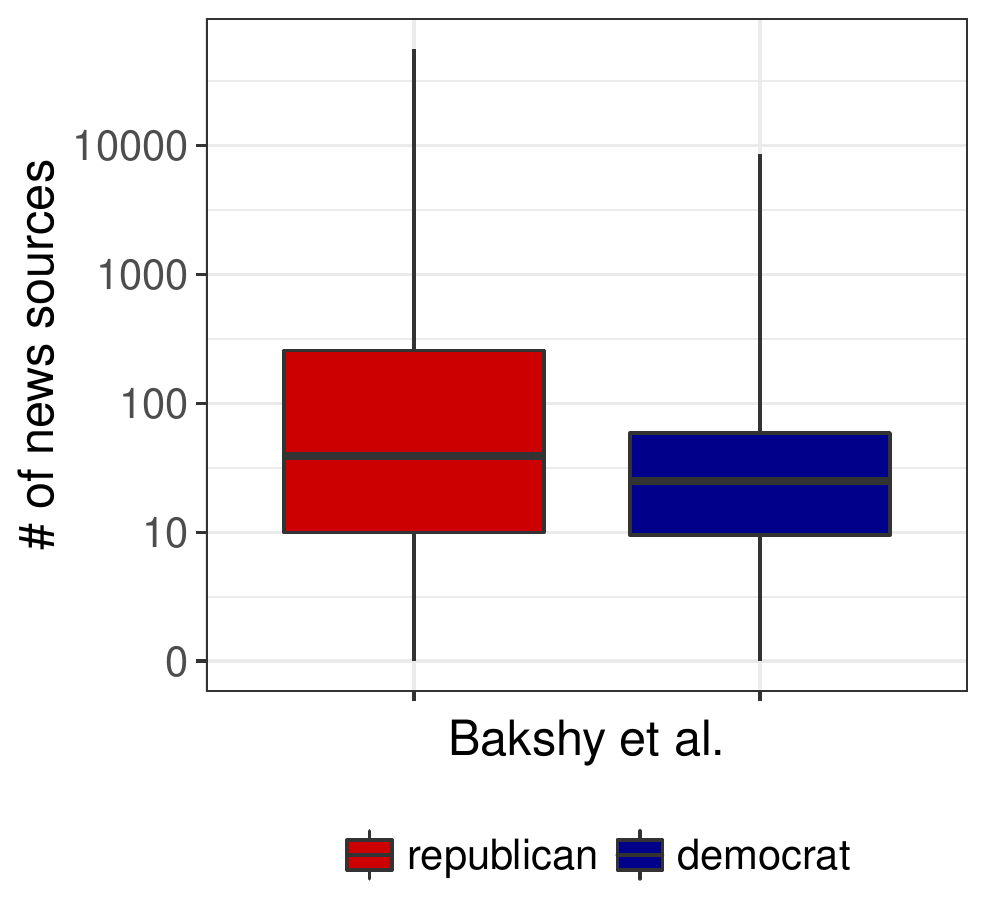}
\caption{}
\label{fig:Bakshy}
\end{subfigure}
\begin{subfigure}{0.24\textwidth}
\includegraphics[width=0.9\linewidth]{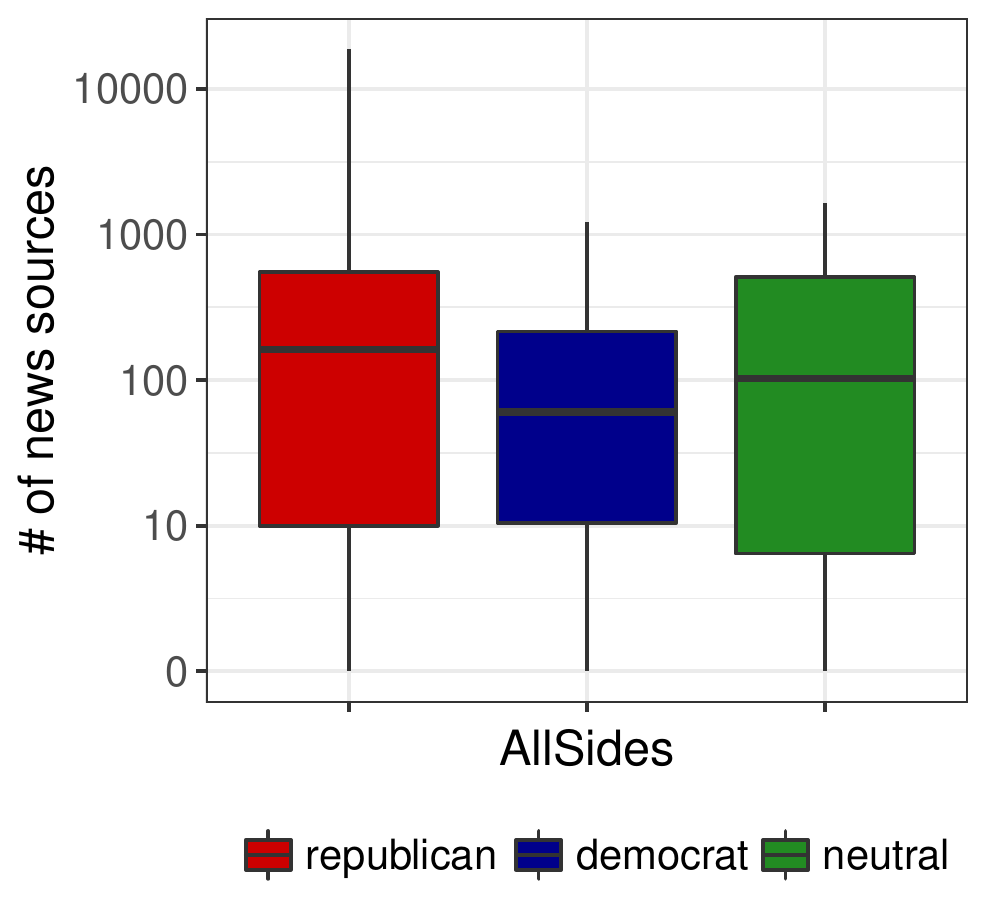}
\caption{}
\label{fig:AllSides}
\end{subfigure}

\caption{\textbf{Number of times news sources were shared in Gab posts categorized as news, grouped by political leaning (Republican, Democrat, and Neutral) as inferred by Budak et al. (a), Mitchell et al. (b),  Bakshy et al. (c), and AllSides (d). Each box plot shows minimum, 25-percentile, median, 75-percentile and maximum.}}\vspace{-15px}
\label{fig:SharedOnGab}
\end{figure*}

Figure~\ref{fig:SharedOnGab} shows box plots of the number of times news sources inferred as Republican, Democrat, and Neutral have been shared on Gab posts categorized as news. Clearly, news outlets biased towards conservative audience are consistently shared on average more than the others across all four methodologies. In particular, considering only news outlets found in the study of Budak et al. (Figure~\ref{fig:Budak}), Gab posts comprise on average $5,894$ right-leaning news sources (median $2,116$) and only 645.6 leftist news sources (median 538).

\subsection{What are the most shared domains?}
First, we extract the domains of all links we collected from $463,663$ Gab posts categorized as news. Table~\ref{tab:domains_top30} shows the proportion of the top 30 most frequently shared domains in this social media\footnote{We exclude from this ranking domains not belonging to news outlets (i.e. twitter.com, i.imgur.com and pbs.twimg.com)}. We observe that \url{ussanews.com} is the most frequently shared news domain, accounting for more than 16\% of the news URLs shared on Gab, followed by \url{youtube.com}, which accounts for more than 15\%. Although YouTube is not specifically designed for news, this social media is widely used for publishing news videos \cite{peer2011youtube, hanson2008youtube}. Moreover, we note a substantial presence of alternative news outlets like \url{breitbart.com}, \url{zerohedge.com}, \url{infowars.com} and \url{thegatewaypundit.com}, which correspond to more than 5\%, 2\%, 1.9\% and 1.5\%, respectively. Finally, the newspaper \url{dailymail.co.uk} (1.5\%) figures among the most shared domains, followed by \url{rt.com} (1.4\%), \url{dailycaller.com} (1.2\%) and \url{fightagainst-tyranny.com} (1.2\%).

\begin{table}[tb]
\resizebox{\columnwidth}{!}{
\centering 
 \begin{tabular} {|l|l|c|r|r|}
 \hline
  \textbf{Source} & \textbf{Domain}  & \textbf{(\%)} \cr \hline
USSA News & ussanews.com & 16.40 \cr \hline	
Youtube & youtube.com  & 15.99  \cr \hline 
Breitbart & breitbart.com & 5.17	\cr \hline 
Zero Hedge & zerohedge.com & 2.08 	\cr	\hline 
Infowars & infowars.com	& 1.91 	\cr \hline 
The Gateway Pundit & thegatewaypundit.com & 1.59 	\cr	\hline 
Daily Mail Online &  dailymail.co.uk & 1.52 	\cr	\hline 
RT News & rt.com & 1.48 	\cr	\hline 
The Daily Caller & dailycaller.com & 1.25 	\cr	\hline 
Tyranny News &  fightagainst-tyranny.com & 1.21 	\cr \hline 
EXPRESS & express.co.uk & 1.20 	\cr	\hline 
Fox News & foxnews.com & 0.99 	\cr	\hline 
Truth Feed News & truthfeed.com & 0.63 	\cr	\hline 
Sputnik News & sputniknews.com & 0.63 	\cr	\hline 
ABC News & abcnews.go.com & 0.57 	\cr	\hline 
Reuters & reuters.com & 0.50 	\cr	\hline 
Washington Examiner & washingtonexaminer.com & 0.49  \cr	\hline
Behoerdenstress Whistleblower & behoerdenstress.de & 0.46 	\cr	\hline 
The Hill & thehill.com & 0.46 	\cr	\hline 
Worldwide Weird News & worldwideweirdnews.com & 0.45 	\cr	\hline 
The Daily Informer & thedailyinformer.net & 0.44 	\cr \hline
WND & wnd.com & 0.44  \cr	\hline
New York Post & nypost.com & 0.41 	\cr	\hline
The Washington Post & washingtonpost.com& 0.34 	\cr	\hline
The Last Refuge & theconservativetreehouse.com & 0.34 	\cr	\hline
The Telegraph & telegraph.co.uk & 0.33 	\cr	\hline
BBC News & bbc.com	& 0.33  \cr	\hline
The Washington Free Beacon & freebeacon.com	& 0.32 	\cr	\hline
INDEPENDENT &  independent.co.uk & 0.30 	\cr	\hline
The Sun & thesun.co.uk	& 0.30 	\cr \hline 
 \end{tabular}}
  \caption{\textbf{Top 30 news sources in posts and their respective domain, percentage over all posts.}}\label{tab:domains_top30}\vspace{-15px}
 \end{table}

 Some news domains widely shared in Gab are not very popular according to Alexa.com. For example, the most shared domain in Gab (\url{ussanews.com}) appears in a very low position in the Alexa rank ($409,260$). The same is true for others domains such as \url{fightagainst-tyranny.com}, \url{truthfeed.com}, \url{behoerdenstress.de}, \url{worldwideweirdnews.com} and \url{thedailyinformer.net}.

Overall, these results show that there is a large diversity of news domains shared in Gab. While most domains are shared very few times, a few domains comprise most of the shared URLs related to news. Interestingly, there is {\bf no intersection} between the top 30 most shared domains in Gab and the top 10 most shared news domains in traditional media like Facebook\footnote{https://www.statista.com/statistics/265830/facebook-daily-newspapers-top-ten/}. On Twitter, previous works showed that news URLs from online newspapers like ``The New York Times'' are shared very often \cite{reis2017demographics}. However, this domain does not appear frequently in Gab shares, highlighting important differences between these social networks.

 \begin{table}[t] \centering 
\vspace*{2mm}
 \begin{tabular} {|l|c|c|}
 \hline
 \textbf{URL News}  & \textbf{Shares} & \textbf{Bit.ly} \cr \hline
http://bit.ly/2qNboDV &	99 & 949 \cr \hline		
https://youtu.be/4Emd7urcWBc &	60 & -	   \cr 	\hline
 https://wikileaks.org/ciav7p1/ &	44 & 4,025	\cr 	\hline
 https://youtu.be/MHZSfhd1X\_8 &	39 & - \cr\hline	
 http://bit.ly/2EJ8WpZ &35 & -	\cr 	\hline
 http://washex.am/2ivMBhv &	35 & 772 \cr		\hline
 http://dailym.ai/2kubLOX &	32 & 0 \cr	\hline	
 http://bit.ly/2nZdrD7 &	27 & - \cr\hline	
 http://dailym.ai/2qSZ5mi &	25 & - \cr		\hline
 http://bit.ly/2nB285x &	24 & 4,492	 \cr 	\hline
 http://bit.ly/2C0QdW1 &	24 & 1 \cr	\hline	
 http://bit.ly/2Es9CAT &	24 & - \cr		\hline
 http://bit.ly/2kSgBZC &	24 & 7,356 \cr	\hline
 http://bit.ly/2iHOmLy &	24 & 274 \cr	\hline
 http://bit.ly/2nZlXlK &	24 & - \cr \hline		
 \end{tabular} 
 \caption{\textbf{Top 15 most popular links shared in Gab, their number of shares in Gab and popularity according to Bit.ly (larger means more popular).}}\label{tab:links_top15_popular-links}\vspace{-15px}
 \end{table}

\subsection{What are the top stories shared in Gab?}

Next, we shift our focus to understand what are the top stories shared in the category news of the platform.

Table~\ref{tab:links_top15_popular-links} shows the 15 most popular links shared in Gab, i.e.\ the  most frequently posted links. We also show their popularity according to Bit.ly\footnote{\url{http://bit.ly/}}, which we use as a proxy for the number of clicks made to each of these links. Bit.ly is a well-known URL shortening service that shortens millions of URLs daily. The service API provides the possibility of checking the total number of clicks that a shortened link has received \cite{antoniades2011we}. In general, these links presented in Table \ref{tab:links_top15_popular-links} have been posted on average 36 times, where the first link in the rank appears in 99 posts and the last one appears in 24 posts. Some popular links shared in Gab have Bit.ly popularity higher than links to popular news sources such as BBC News, DailyMail, New York Times, and Reuters \cite{reis2015breaking}.

The most popular link is a petition created on May 19, 2017, requesting action of the Congress regarding the following issue: ``Appoint a Special Prosecutor to investigate the murder of Seth Rich, the alleged Wikileaks email leaker". Other popular links include YouTube videos (the second most popular link talks about the alternative media), WikiLeaks news, Donald Trump-related news, and also news which might not be completely accurate. For instance, the sixth most popular link has been rated by the fact-checking website Snopes as mostly false\footnote{\url{https://www.snopes.com/child-prostitution-legalized-in-california/}}. This suggests that Gab is prone to the dissemination of fake news, similarly to other social networks. However, the fake news that spread in Gab are usually those that reinforce right-leaning beliefs.

\section{Conclusion} \label{sec:conclusion}

In this paper we characterize Gab, a social network that emerged advocating liberty and freedom of expression, but received several criticisms about the content shared in it. As the debate between hate speech and free speech is an open and very controversial issue, Gab is an evident source of data for many researchers to explore deeper all those concerns. 
In this direction, our study provides a characterization of users and content shared on Gab and also contributes to understanding the behavior of such users without the strict policies of moderation from other medias. 

Our findings show that Gab is a very politically oriented system that hosts known banned users from other social networks. We also show that the majority of Gab users are conservative, male, and Caucasian. Gab is also crowded by extremist users. Our analysis of what users share in Gab unveiled a lot of political statements. Posts indicate that, while users support free speech, a small part of the posts not only mirror political views but incorporate hate speech. 

Our analyses also show the existence of a great variety of news domains shared within Gab. Most of these domains are not popular in other social media or on the Internet as a whole, and part of them are known for spreading politics-related news. These results indicate that an unmoderated social media such as Gab has become an echo chamber for right-leaning content dissemination.


\section*{Acknowledgments}
This research was partly supported by CAPES, MASWeb (grant FAPEMIG/PRONEX APQ-01400-14), Fapemig, and Humboldt Foundation. We also thank Juhi Kulshrestha for measuring the political leaning of the users in Gab. 

\bibliographystyle{IEEEtran}
\small{\bibliography{Bibliography-File}}

\end{document}